# Multi-LLM Consensus Framework for Evaluating Banking-Sector NIDS Dataset Coverage of MITRE ATT&CK Techniques

Sanjida Khanom[1], Sadia Afrin Khan[1], Adrita Rahman Tory[1], Md. Ahsan Habib[1], and Khondokar Fida Hasan[2]⋆

Bangladesh University of Business and Technology (BUBT), Mirpur-2, Dhaka-1216, Bangladesh
University of New South Wales (UNSW), ACT 2601, Australia
fida.hasan@unsw.edu.au

**Abstract.** The systemic criticality of global banking networks has rendered them high-priority targets for advanced persistent threats, necessitating Network Intrusion Detection Systems (NIDS) whose operational effectiveness must extend beyond statistical accuracy. However, a significant validation gap persists between experimental NIDS performance and real-world effectiveness: NIDS models that achieve high accuracy on standard benchmarks often fail in operational banking environments because generic datasets lack sector-specific patterns, such as SWIFT and ATM-related intrusions, that characterize real financial threats. To address this, the paper investigates a sector-aware evaluation methodology that systematically assesses how well existing NIDS benchmark datasets cover the attack behaviors most relevant to banking infrastructure. The methodology maps documented adversary behaviors from the MITRE ATT&CK knowledge base to NIDS benchmarks while enforcing the realistic sensor limitations defined by NIST SP 800-94. Leveraging a multi-LLM consensus engine with four state-of-the-art models, we evaluated 210 banking-specific adversary techniques to derive a baseline of 68 network-observable behaviors for systematic coverage analysis. Results across five benchmark datasets demonstrate that UNSW-NB15 achieves the highest utility with an 82.2% weighted coverage score (though only 18.4% reflects direct, technique-level evidence), while CIC-DDoS2019 reveals an 89.9% blind spot for core banking behaviors. These findings establish a reproducible foundation for sector-aware NIDS evaluation and highlight the urgent need for banking-native datasets.

**Keywords:** Network Intrusion Detection System (NIDS) · Banking Security · Cybersecurity · MITRE ATT&CK · Large Language Models.

## 1 Introduction

Modern banking infrastructure constitutes the backbone of global financial stability, facilitating high-value transactions through interconnected networks in-cluding SWIFT, real-time gross settlement (RTGS) systems, and automated teller machine (ATM) grids [4,7]. This digital transformation, while enhancing operational efficiency, has simultaneously expanded the attack surface for sophisticated adversaries. Recent campaigns by groups such as APT38, Lazarus, and Carbanak demonstrate that banking networks are no longer opportunistic targets but strategic objectives for state-sponsored and financially motivated threat actors [1,16]. These adversaries employ multi-stage intrusions specifically designed to exploit the unique protocols and transaction flows inherent to financial market infrastructures (FMIs), rendering generic cybersecurity measures insufficient [16].

⋆ Corresponding author, *Email: fida.hasan@unsw.edu.au*

To defend these critical environments, Network Intrusion Detection Systems (NIDS) are widely deployed as passive sensors monitoring traffic flows for unauthorized or anomalous activities. NIST identifies NIDS as a core mechanism for continuous network monitoring and policy enforcement under SP 800-94 guidelines [21]. However, the effectiveness of these systems is fundamentally constrained by the representativeness of their training and evaluation datasets [19,13]. Despite this well-documented limitation, ML-based NIDS research continues to rely on a small set of general-purpose benchmarks without critically evaluating their suitability for sector-specific deployment [13,20].

A growing body of evidence suggests that NIDS models achieving high accuracy on standard benchmarks often underperform when deployed in operational banking environments, primarily because experimental datasets lack the sector-specific attack patterns such as transaction abuse, credential misuse, and interbank communication manipulation that characterize real-world financial threats [15,20]. This validation gap stems from a fundamental mismatch between available benchmarks and operational requirements. Popular datasets including NSL-KDD, UNSW-NB15, and CICIDS2017 were designed for generic enterprise IT environments, focusing primarily on coarse-grained attacks such as denial-of-service, port scanning, and brute-force activities [6,17,23]. While suitable for baseline algorithmic evaluation, these datasets provide limited coverage of banking-specific adversarial behaviors, particularly those targeting payment protocols or exhibiting the temporal continuity required for multi-stage intrusion detection [19]. Consequently, performance metrics derived from these benchmarks risk creating false assurance regarding operational resilience [24].

Recent studies have adopted the MITRE ATT&CK framework as a structured ontology for assessing detection coverage. ATT&CK models real-world adversary behavior through documented tactics and techniques, enabling systematic reasoning about which attack behaviors are observable within network traffic [15,14,12,10]. Some efforts have extended this to specific domains: Bagui et al. [3] developed a network traffic dataset explicitly mapped to ATT&CK techniques, while Wu et al. [26] demonstrated the applicability of Large Language Models(LLM) to automated threat modeling within banking systems. However, existing mapping approaches suffer from two critical deficiencies. First, they typically treat all ATT&CK techniques as equally detectable by network sensors, while ignoring any standard constraints on encryption and endpoint visibility[21,19].

Second, manual mapping methodologies do not scale to the granularity required for high-stakes banking environments [8].
This paper proposes a sector-aware evaluation methodology that bridges the gap between experimental NIDS validation and operational banking security requirements. The key contributions of this work are listed as follows:

- We introduce a banking-sector evaluation framework that filters MITRE ATT&CK techniques through NIST SP 800-94 passive NIDS visibility limits (passive-only sensing, no endpoint data, no TLS decryption) before mapping to datasets, preventing inflated "coverage" claims and defining a realistic observability bound.
- We build a reproducible baseline of banking-relevant ATT&CK techniques (core banking, SWIFT/RTGS, ATM environments) and curate the subset that is network-observable under passive constraints, providing a standard reference for sector-aware dataset validation.
- We propose a consensus-driven technique-to-dataset mapping procedure and a priority-weighted coverage metric that scores the utility of the data set by emphasizing the techniques most important to banking adversaries, shifting evaluation from raw counts to representativeness relevant to the threat.

The remainder of this paper is organized as follows. Section 2 reviews related work. Section 3 presents the proposed methodology. Section 4 reports the experimental results. Section 5 discusses the implications and limitations of the findings. Section 6 concludes.

## 2 Background and Related Works

### 2.1 Background

The digital transformation of the banking sector has necessitated highly interconnected infrastructures involving cloud services and global networks like SWIFT [4,7,16]. To mitigate growing threats including ransomware and lateral movement [1,16], banks deploy Network Intrusion Detection Systems (NIDS) across both external gateways and internal segments to monitor traffic patterns and privilege misuse in accordance with National Institute of Standards and Technology (NIST) standards [21,19]. Figure 1 illustrates a representative passive NIDS deployment architecture within critical banking infrastructure, structured around a two-firewall DMZ (demilitarized zone) model consistent with NIST SP 800-94 sensor placement guidelines [21]. Three passive tap sensors monitor traffic at key chokepoints: at the internet gateway boundary, at the SWIFT interbank connection, and within the internal core banking segment covering transaction flows between database clusters, payment engines, and ATM controllers [21]. Standard benchmarks such as KDD Cup 99, NSL-KDD [23], UNSW-NB15, and CICIDS2017 are widely used due to their accessibility, yet they primarily reflect generic enterprise IT environments and coarse-grained attack vectors such as denial-of-service, port scanning, and brute-force activity [6,17,23,11]. A growing body of evidence highlights that public NIDS datasets cover only a fraction of known attack behaviors [13], and that datasets lacking transaction-specific context and financial communication protocols provide limited basis for evaluating NIDS deployed within banking infrastructure [19,20,24]. The MITRE ATT&CK framework provides a structured ontology for reasoning about these gaps, modelling real-world adversary behavior through documented tactics and techniques to enable systematic assessment of which attack behaviors are observable within network traffic [15,14].

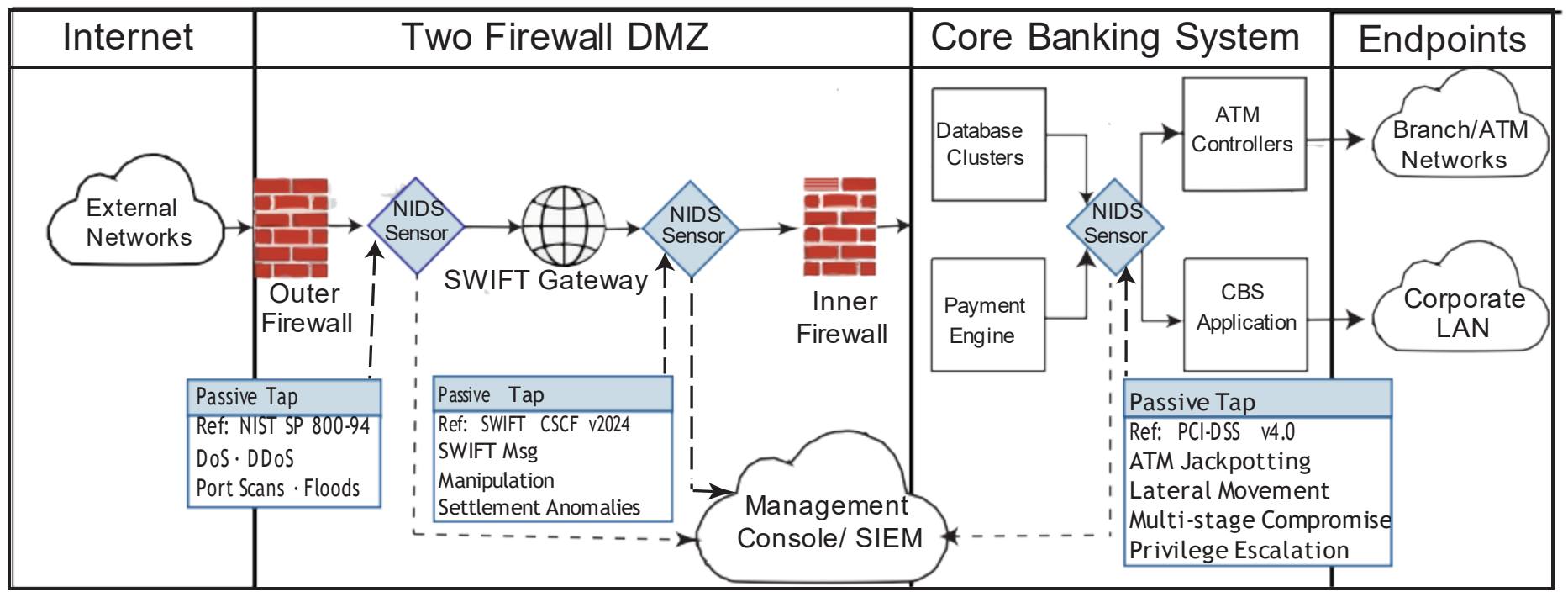

**Fig. 1.** Passive NIDS Deployment Architecture in Banking Infrastructure

### 2.2 Related Works

Building on these identified gaps, recent research has sought more structured approaches for assessing detection coverage. Table 1 summarizes key related works and the gaps this paper addresses.

As Table 1 illustrates, existing ATT&CK-based mapping approaches suffer from two critical deficiencies. First, they typically treat all ATT&CK techniques as equally detectable by network sensors, ignoring NIST SP 800-94 constraints on encryption, endpoint visibility, and passive monitoring limitations [21,19]. Second, manual mapping methodologies do not scale to the granularity required for high-stakes banking environments [8].

While LLMs are increasingly used for security tasks like threat modeling, individual model outputs are often susceptible to reasoning biases [8,26]. To ensure high-stakes reliability in banking environments, recent studies propose consensus-based frameworks [18,24]. The most closely related prior work, Tory et al. [24], conducted a gap analysis of NIDS datasets for the energy sector using a dual-LLM validation approach, but did not apply NIST SP 800-94 passive constraints or address banking-specific threat actors. Furthermore, existing consensus and validation frameworks [8,18,24] lack banking-specific threat modeling to establish a reproducible baseline of network-observable techniques. This paper addresses these gaps by introducing a NIST-constrained detectability framework that filters ATT&CK techniques based on passive NIDS observability. The resulting baseline provides a realistic upper bound for sector-aware evaluation of banking NIDS datasets.

**Table 1.** Qualitative Comparison of Related Works in NIDS Datasets and MITRE ATT&CK Mapping for Banking Sector

| **Ref** | **Scope** | **Methodology / Focus** | **Identified Gaps** | **This Paper's Response** |
|---|---|---|---|---|
| Hasan & Rahman Tory [19] | Generic NIDS | MITRE ATT&CK -based evaluation of NIDS datasets | Missing NIST SP 800-94 constraints and banking-specific context | Applies strict NIST SP 800-94 passive monitoring limits (no TLS/endpoint visibility) tailored for banking environments |
| Kinnunen [14] | Cyber Defense | MITRE ATT&CK threat detection gap analysis | Manual and unscalable mapping; lacks financial-sector specificity | Introduces multi-LLM automated mapping with banking-aware filtering |
| Bagui et al. [3] | Cloud IT | ATT&CK -aligned network traffic dataset construction | Lacks financial transaction context (e.g., SWIFT, RTGS) | Extends baseline with 210 banking-relevant techniques (ATM, SWIFT operations) |
| Daniel et al. [8] | NIDS Labeling | Comparison of ML vs. LLMs for ATT&CK rule labeling | Single-LLM bias; absence of consensus-based validation | Proposes 4-model LLM consensus with a 75% agreement threshold |
| Wu et al. [26] | Banking | Automated LLM-driven threat modeling for banking systems | Overestimated coverage due to ignoring passive NIDS constraints | Introduces 3-tier NIDS filtering aligned with NIST SP 800-94 limitations |
| Tory et al. [24] | Energy Sector (IT/OT) | ATT&CK gap analysis for energy NIDS datasets with dual-LLM validation (Claude + Gemini) | No banking focus; limited to 2-LLM validation; lacks NIST SP 800-94 passive constraints | Extends methodology to banking using 4-model consensus and NIST SP 800-94-compliant filtering for SWIFT/ATM threat scenarios |

## 3 Methodology

The evaluation framework operates through a three-stage pipeline: (1) banking-sector threat identification and technique extraction, (2) NIST-constrained detectability classification, and (3) multi-LLM consensus mapping against benchmark datasets. Figure 2 illustrates this workflow.

### 3.1 Scope Definition and Threat Identification

Although the MITRE ATT&CK framework classifies financial entities under a unified "Financial Services" sector, this study employs a narrower definition to isolate threats specifically targeting core banking infrastructure. Banking institutions are defined here as entities operating critical financial market infrastructures (FMIs), specifically interbank payment systems (such as SWIFT and RTGS), automated teller machine (ATM) networks, and core banking systems facilitating deposit and settlement functions [4]. This distinction excludes broader financial entities like insurance providers, investment firms, and cryptocurrency exchanges, which do not operate the same systemically important payment infrastructure [7].

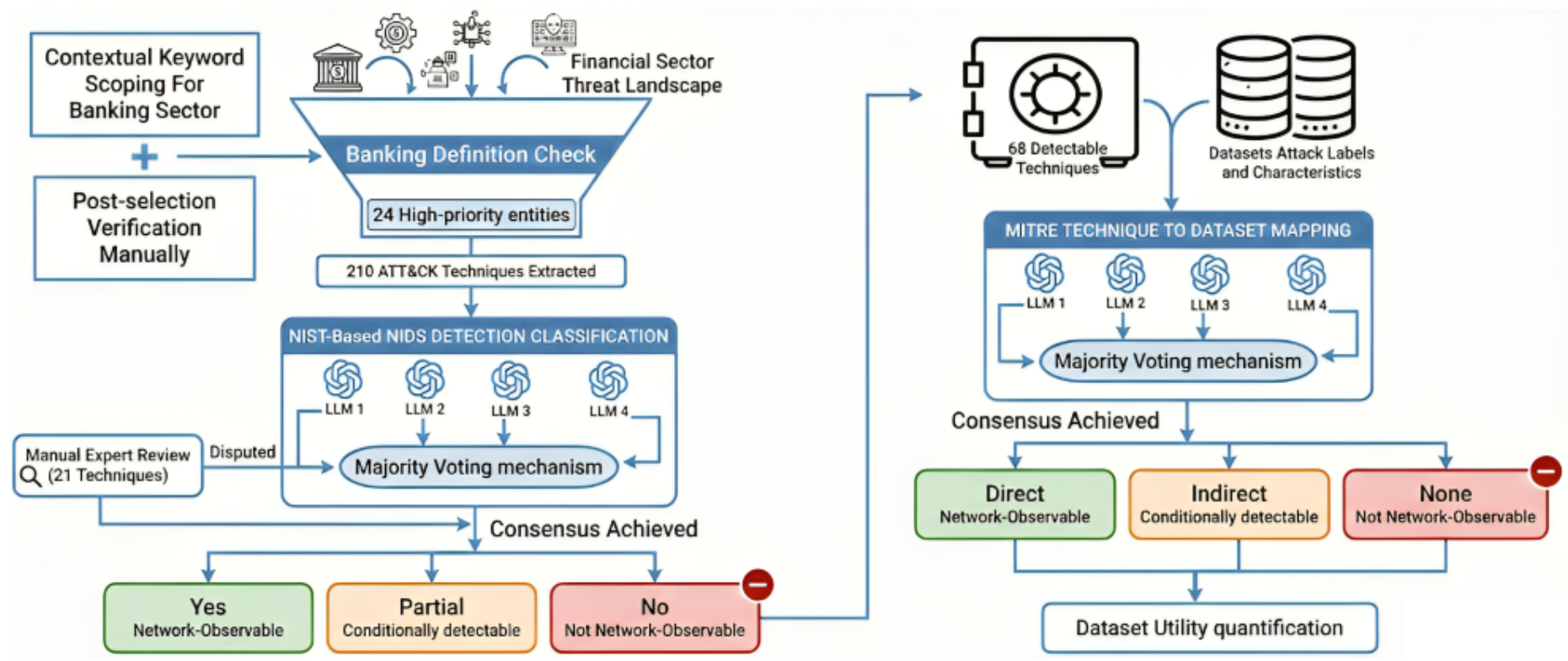

**Fig. 2.** Multi-LLM Consensus Framework for Banking-Sector NIDS Dataset Evaluation. The pipeline progresses from entity extraction (210 techniques) through NIST-based detectability filtering (yielding 68 network-observable techniques) to dataset utility quantification via majority voting consensus.

Leveraging the MITRE ATT&CK framework (v18) [15], we used a keyword-based filtering method, similar to [19], to identify potential banking-sector adversaries. Candidates were manually verified to ensure their targets matched our banking definition, excluding groups focused only on peripheral financial services. This yielded 12 software families (Dridex, TrickBot, QakBot, Dyre, Carbanak, Carberp, Ursnif, IcedID, Bankshot, Zeus Panda, Wiper, and TrickMo) and 12 threat groups (Indrik Spider, APT38, Lazarus Group, Cobalt Group, Andariel, GCMAN, Malteiro, Silence, TA505, DarkVishnya, RTM, and Carbanak). Notably, Carbanak appears both as a malware entry (S0030) and as a threat group (G0008), as they are distinct entities in MITRE ATT&CK. Technique extraction was performed programmatically via Python against the MITRE ATT&CK Enterprise STIX/JSON repository in November 2025, using version 18.1 (released October 28, 2025) [15], resulting in 210 unique techniques.

**Table 2.** Passive NIDS Sensor Constraints per NIST SP 800-94

| Constraint | Description | NIST Reference |
|---|---|---|
| **No TLS decryption** | Encrypted payloads (HTTPS, SSH, VPN) are opaque to the sensor; only metadata (e.g., SNI, certificates) remains visible. | [21] Section 4.3.3.4 |
| **No endpoint visibility** | Internal process execution, filesystem changes, memory states, and local authentication are outside the sensor's scope. | [21] Sections 4.5, 7.1 |
| **Passive-only monitoring** | Sensor observes traffic copies via SPAN ports or TAPs; it lacks inline prevention or active response capabilities. | [21] Section 4.2 |

### 3.2 Detectability Classification Framework

To categorize techniques by passive NIDS detectability, we developed a classification framework operationalizing NIST SP 800-94 capabilities [21]. The model assumes passive sensor deployment via network taps, subject to three explicit constraints summarized in Table 2.
Four LLMs (Claude Opus 4.5, GPT-5.2, Grok 4.1, and Gemini 3 Pro) independently evaluated each of the 210 techniques. To ensure cross-model consistency, the prompting strategy applied six prompt engineering principles summarized in Table 3; generalized complete prompt templates are provided in Appendix A.

**Table 3.** Prompt Engineering Techniques Applied for Cross-Model Consistency

| Technique | Implementation | Rationale for Consistency |
|---|---|---|
| **Structural tags** | XML-style delimiters (<task>, <constraints>) | Creates unambiguous boundaries; prevents models from conflating instructions with examples [2]. |
| **Role assignment** | "Network security analyst" (Stage 1); "Senior CTI Analyst" (Stage 2). | Anchors domain-specific vocabulary and reasoning patterns uniformly across models. |
| **Constraint framing** | Fixed declarative parameters: "The sensor cannot..." vs "typically doesn't." | Eliminates interpretation latitude; reduces variance in ambiguous edge cases. |
| **Positive instructions** | "Assign exactly one label" instead of "do not assign multiple labels." | Outperforms negative framing by providing a clearer, direct action directive [5]. |
| **Explicit output limits** | "Maximum 15 words" specified for reasoning columns. | Quantified limits ensure uniform output; qualitative instructions vary too widely by model. |
| **Few-shot calibration** | 8 examples (YES, PARTIAL, NO) embedded in Stage 1 (see Appendix A). | Calibrates decision boundaries; reduces model-specific classification drift. |

Each LLM assigned one of three classification labels defined in Table 4, operationalizing NIST SP 800-94 detection gradients.

**Table 4.** NIDS Detectability Classification Scheme

| Label | Definition | NIST Basis |
|---|---|---|
| **YES** | Core behavior produces observable network artifacts, including signatures, anomalies, or protocol violations. | Event types per [21] Section 4.3.3.1 |
| **PARTIAL** | Detection conditional on encryption state, relies on metadata inference, or varies by implementation. | Flow analysis per [21] Section 6.2 |
| **NO** | Executes locally without a network footprint, or traffic is indistinguishable from legitimate activity. | Host-based IDPS requirement per [21] Section 7.1 |

Consensus required agreement from at least three of four LLMs (75%). This threshold balances inclusivity and rigor: while unanimous consensus (100%) eliminates false positives, it risks excluding valid techniques due to single-model bias;

conversely, simple majority (50%) permits individual model errors to dictate classification. The 75% threshold ensures robust agreement while accommodating legitimate interpretive variation regarding network artifact observability [18]. Of 210 techniques, 21 (10.0%) failed to achieve consensus and required manual adjudication, comprising 19 cases of 2-2 splits and 2 cases of three-way disagreement (2-1-1 pattern). Disputed classifications were resolved through structured expert review. Conservative classification was prioritized; techniques with ambiguous network observability were classified as PARTIAL rather than YES to avoid inflating detectability estimates. This process identified 68 techniques as NIDS-detectable, forming the scope for dataset mapping.

### 3.3 Dataset Mapping and Utility Metrics

**Table 5.** Dataset Mapping Classification

| Label | Definition |
|---|---|
| **Direct** | Dataset explicitly contains attack traffic that directly implements the technique's core behavior with observable network artifacts. |
| **Indirect** | Dataset contains related network activity that partially represents the technique or captures associated artifacts without full behavioral alignment. |
| **None** | No dataset labels correspond to the technique, indicating the attack vector is absent from the dataset's scope. |

To evaluate coverage, we developed structured dataset profiles characterizing each benchmark's attack taxonomy, network artifacts, and known limitations. Five NIDS datasets were analyzed: CICIDS2017[6], CIC-DDoS2019[22], UNSW-NB15[17], UWF-ZeekData22[25], and CTU-13[9].

**Dataset Profiling:** Each profile documented: (1) attack categories and their distribution, (2) observable network artifacts including ports, protocols, flags, and payload characteristics, (3) dataset-specific features relevant to detection, and (4) known limitations affecting technique coverage, such as data format constraints, class imbalance, or temporal obsolescence.

**Table 6.** Dispute Resolution Rules for 2-2 Mapping Splits

| Dispute Pattern | Resolution | Rationale |
|---|---|---|
| Direct vs. Indirect | **Indirect** | Conservative approach; acknowledges partial coverage without overstating behavioral alignment. |
| Direct vs. None | **Indirect** | Recognizes potential coverage while accounting for significant classification uncertainty. |
| Indirect vs. None | **Excluded** | Insufficient consensus to claim coverage; removed to prevent speculative alignment. |

**Semantic Mapping Process:** The same four LLMs independently evaluated each technique against the dataset profiles. Each model received the 68-technique master list, a structured dataset profile, and reasoning instructions demonstrating the mapping logic. Each mapping was assigned one of three classification labels defined in Table 5.
Mapping consensus required 75% agreement, consistent with Stage 1. Unanimous (4/4) and consensus (3/4) classifications were accepted directly. Disputed cases (2-2 splits) were resolved through predefined conservative rules summarized in Table 6, ensuring reproducibility and eliminating subjective bias.
Across all datasets, 12 techniques (17.6% of the 68-technique baseline) were excluded due to Indirect-None disputes, ensuring reported coverage represents high-confidence mappings only.

**Priority Scoring:** Each technique received a composite priority score (0-100). To rank the criticality of each technique, a composite Priority Score (PS), ranging from 0–100, is calculated as follows:

$$PS = U_e + I_t + D_g \quad (1)$$

Where:

- $U_e \in [0, 40]$ represents the Entity Usage (frequency of use in banking sector attacks).
- $I_t \in [0, 40]$ represents the Tactic Importance (criticality of the associated MITRE tactic).
- $D_g \in [0, 20]$ represents the Group Diversity (number of distinct threat actors employing the technique).

Entity Usage reflects threat actor and malware adoption breadth. Tactic Importance derives from empirical tactic distribution across the banking threat corpus. Group Diversity rewards techniques employed by multiple distinct threat groups. The 40/40/20 weighting prioritizes direct threat adoption and operational impact. Techniques were categorized as High-Priority (≥6 entities), Common (4-5), Medium (2-3), or Rare (1).

**Coverage Metrics:** Dataset utility was quantified using weighted coverage:

$$\text{Coverage (Weighted)} = \frac{\sum_{i=1}^{n} (w_i \cdot c_i)}{\sum_{i=1}^{n} w_i} \times 100 \quad (2)$$

Where $w_i$ is the technique's priority score, and $c_i$ is binary (1 if Direct/Indirect, 0 if None), ensuring high-priority techniques contribute proportionally to the assessment.

### 3.4 Reproducibility and Artifact Availability

All primary artifacts are publicly archived at https://github.com/adritatori/BankSectorNIDSdatasetCoverage, organized into three directories corresponding to the three methodological stages. The consensus engine comprised GPT-5.2 (OpenAI), Claude Opus 4.5 (Anthropic), Gemini 3 Pro (Google), and Grok 4.1 (xAI), accessed via standard consumer chat interfaces between November 2025 and January 2026. Because inference parameters such as temperature and top-p were not user-configurable through consumer interfaces, outputs are reproducible in structure but may exhibit minor variation due to non-deterministic sampling[18,24]. The ATT&CK framework version used throughout is v18, accessed via https://attack.mitre.org/ in November 2025. Prompt templates and few-shot calibration examples are provided in Appendix A.

# 4 Results

The evaluation focuses on the representation of NIDS-detectable MITRE ATT&CK techniques associated with adversaries targeting core banking infrastructure.

## 4.1 NIDS Detectability Classification of Banking Sector Techniques

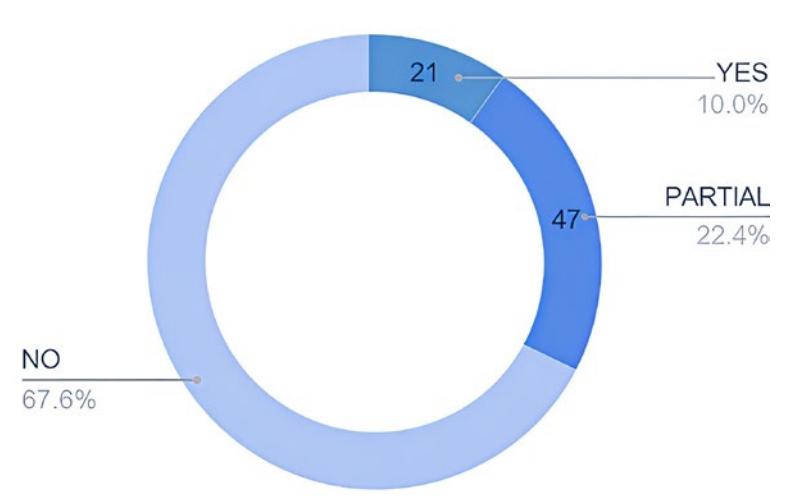

**Fig. 3.** NIDS Detectability Classification by LLM Evaluator.

From the initial set of 210 banking-sector techniques, 68 (32.4%) were classified as NIDS-detectable (YES or PARTIAL) under strict passive NIDS constraints. The remaining 142 (67.6%) were classified as NO, primarily due to host-resident execution or encrypted traffic indistinguishable from legitimate activity. Figure 3 illustrates the three-way classification distribution across all 210 techniques, where 21 (10.0%) received YES and 47 (22.4%) received PARTIAL labels. The combined 68 YES and PARTIAL techniques (32.4%) constitute the NIDS-detectable baseline, establishing the realistic upper bound for passive NIDS evaluation within the banking threat model.

## 4.2 Inter-Model Agreement on Dataset Mapping

Consensus outcomes for mapping the 68 detectable techniques to benchmark datasets are summarized in Table 7. CIC-DDoS2019 exhibits the strongest agreement (92.6% unanimous or consensus classifications), attributable to its narrow, well-defined attack scope. Conversely, CTU-13 demonstrates the highest ambiguity, with 50.0% of techniques resulting in disputed (2-2) classifications and zero instances of weak agreement. This reflects CTU-13's coarse-grained, campaign-oriented labeling structure, which complicates fine-grained attribution of techniques compared to datasets with explicit attack labels.
CIC-DDoS2019 is mainly focused on DDoS attacks which generate clear and repetitive traffic patterns. This makes detection easier and leads to higher agreement among techniques [22]. In contrast, multi-attack datasets include diverse and complex attack behaviors resulting in lower consistency. Therefore, the high performance of CIC-DDoS2019 reflects DDoS-specific detection capability rather than general intrusion detection performance. This limitation has been acknowledged to ensure fair comparison and proper interpretation of results.

**Table 7.** Inter-Model Agreement Summary

| Dataset | Total Tech. | Unanimous (4/4) | Consensus (3/4) | Weak (2/4) | Disputed (2-2) |
|---|---|---|---|---|---|
| CICIDS2017 | 68 | 16 (23.5%) | 32 (47.1%) | 8 (11.8%) | 12 (17.6%) |
| CIC-DDoS2019 | 68 | 44 (64.7%) | 19 (27.9%) | 3 (4.4%) | 2 (2.9%) |
| UNSW-NB15 | 68 | 25 (36.8%) | 23 (33.8%) | 9 (13.2%) | 11 (16.2%) |
| UWF-ZeekData22 | 68 | 24 (35.3%) | 18 (26.5%) | 17 (25.0%) | 9 (13.2%) |
| CTU 13 | 68 | 18 (26.4%) | 16 (23.5%) | 0 (0.0%) | 34 (50.0%) |

## 4.3 Dataset Coverage Analysis

The 68 NIDS-detectable techniques were mapped against five benchmark datasets to evaluate the representation of banking-sector threats. Figure 4 summarizes the weighted coverage scores across the evaluated datasets. UNSW-NB15 provides the most comprehensive representation of banking-relevant behaviors, achieving 82.2% total coverage despite limited direct alignment (18.4%). Conversely, CTU-13 exhibits a pronounced imbalance, with substantial indirect representation (40.4%) but minimal direct coverage (9.0%). Finally, CIC-DDoS2019's 89.9% coverage gap reflects its specialized focus on volumetric denial-of-service attacks, rendering it unsuitable for a comprehensive banking-sector security evaluation. Critical techniques exhibit uneven coverage across all evaluated datasets. While exfiltration methods (T1041, T1048) appear in multiple benchmarks, credential access (T1078) and persistence mechanisms (T1053) remain significantly underrepresented. Notably, no dataset provides direct coverage of T1071 (Application Layer Protocol), despite its criticality for Command and Control (C2) detection. Figure 5 illustrates these tactic-level coverage patterns. Direct coverage remains below 18.9% across all datasets, indicating that explicit banking-specific attack implementations are rare. Discovery and Exfiltration emerge as the best-covered tactics, while Initial Access and Impact remain systematically underrepresented within current NIDS benchmarks.

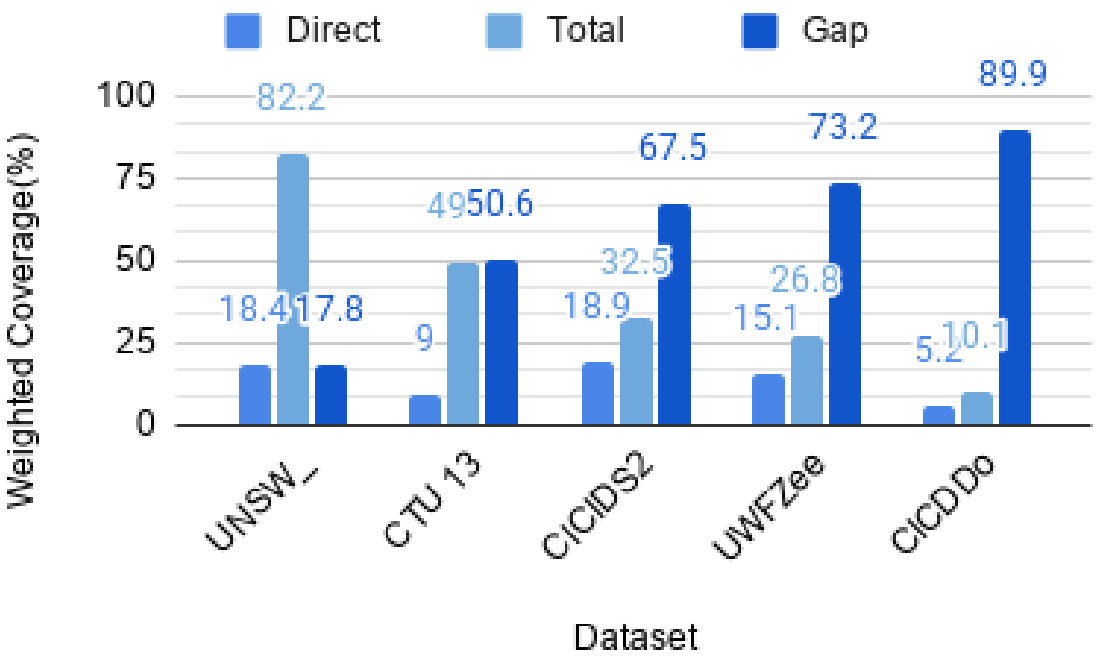

**Fig. 4.** Dataset Coverage Analysis.

## 4.4 Analysis of Mapping Disputes

Disputed dataset mappings reveal systematic ambiguity in ATT&CK technique boundaries. Analysis of the 34 disputed (2–2 split) cases in CTU-13 reveals that disagreements primarily stem from coarse-grained, campaign-level labeling. For the 12 disputed cases in CICIDS2017 and 11 in UNSW-NB15, disagreement typically involved techniques with overlapping network indicators (e.g., distinguishing T1071 from T1041 when both utilize HTTP). In these instances, the conservative default to an Indirect classification prevents false precision while acknowledging potential coverage. The exclusion of 12 techniques (17.6% of the baseline) due to irreconcilable Indirect versus None disputes primarily affected datasets with limited attack diversity, such as CIC-DDoS2019 and UWF-ZeekData22. These exclusions ensure that the reported coverage metrics represent high-confidence mappings rather than speculative alignments.

| | CICDDos2019 | CICIDS2017 | CTU 13 | UNSW_NB15 | UWFZeekData22 |
|---|---|---|---|---|---|
| command-and-control | 17.8 | 49.2 | 58.9 | 100.0 | 0.0 |
| credential-access | 17.6 | 46.7 | 0.0 | 81.8 | 81.8 |
| defense-evasion | 0.0 | 0.0 | 60.1 | 37.8 | 11.6 |
| discovery | 21.5 | 38.5 | 66.8 | 100.0 | 87.5 |
| execution | 0.0 | 56.5 | 50.0 | 56.5 | 0.0 |
| exfiltration | 0.0 | 83.7 | 100.0 | 100.0 | 24.6 |
| impact | 0.0 | 0.0 | 0.0 | 100.0 | 0.0 |
| initial-access | 0.0 | 8.7 | 0.0 | 100.0 | 27.5 |
| lateral-movement | 10.7 | 33.3 | 0.0 | 100.0 | 81.7 |
| persistence | 0.0 | 0.0 | 0.0 | 100.0 | 100.0 |

**Fig. 5.** Tactic-Level ATT&CK Coverage Across Benchmark Datasets

## 5 Discussion

Research findings indicate a substantial validation gap between experimental Network Intrusion Detection Systems (NIDS) and operational banking requirements. Although the UNSW-NB15 dataset covers 82.2% of banking-relevant techniques, its heavy reliance on indirect artifacts (63.8%) suggests it is better suited for anomaly detection than precise technique classification. Systems trained on these datasets may identify suspicious activity without recognizing specific adversary tradecraft, such as ATM jackpotting or SWIFT manipulation. The poor performance of CIC-DDoS2019, which provides only 10.1% coverage, underscores the risks of relying on single-vector datasets. Financial institutions relying on such benchmarks remain vulnerable to 89.9% of the established technique baseline, creating a "false assurance" in NIDS procurement. The limited visibility into Initial Access and Impact tactics confirms that NIDS must be integrated with host-based detection and TLS inspection to ensure comprehensive security.
The multi-LLM consensus methodology mitigates individual model biases. A 75% consensus threshold was selected to balance precision and recall: a lower 50% threshold would expand the detected techniques by 12% at the cost of classification confidence, while a unanimous requirement would have excluded 35% of the baseline. Temporal validity is another concern, as datasets such as CICIDS2017 and UNSW-NB15 predate modern cloud-native banking and TLS 1.3 encryption. The exclusion of 12 techniques (17.6%) likely underestimates actual coverage, as disputed techniques, typically involve use of valid credentials indistinguishable from benign traffic without host-level telemetry. While this study focuses on banking, the results may not fully generalize to cryptocurrency or insurance sectors, which face distinct threat landscapes.
Priority should be given to developing banking-specific benchmarks that include SWIFT message flows and ATM protocols with explicit ATT&CK labeling, enabling supervised learning models to recognize specific adversary behaviors rather than merely detect anomalies. The current 68-technique baseline requires continuous updates to align with evolving MITRE ATT&CK versions and emerging financial threat intelligence. Future research should also explore encrypted traffic analysis to regain visibility into Defense Evasion tactics without requiring full TLS decryption. Extending this methodology to host-based systems would facilitate a more holistic evaluation of multi-layered security architectures across other critical sectors like healthcare and energy.

## 6 Conclusion

Modern banking networks require NIDS evaluation that reflects operationally realistic adversary behavior rather than generic benchmark performance. This paper addresses the validation gap in banking-sector NIDS research by introducing a sector-aware framework that bridges threat intelligence and realistic sensor capabilities. By applying NIST SP 800-94 constraints to the MITRE ATT&CK matrix, the methodology filters 210 banking-relevant techniques into a reproducible baseline of 68 network-observable behaviors. Empirical evaluation of five benchmark datasets reveals that while UNSW-NB15 provides the highest utility at 82.2% coverage, others, such as CIC-DDoS2019, leave an 89.9% blind spot for core banking behaviors. These findings, enabled by a scalable Multi-LLM consensus approach, underscore the need to align dataset selection with sector-specific threat models to avoid operational failures. This framework provides a foundation for future banking-centric benchmarks incorporating SWIFT and ATM protocols and establishes a methodology applicable to security validation across all critical infrastructure domains. Limitations include passive-only visibility assumptions and dependence on model consensus, motivating future work on updated baselines, encrypted-traffic inference, and the development of banking-native datasets incorporating SWIFT/ATM behaviors with explicit ATT&CK labeling.

## Appendix A: LLM Evaluation Prompt Templates

This appendix reproduces the two prompt templates used across all LLM evaluations. Stage 1 was applied once per model (4 runs); Stage 2 once per dataset-model combination (20 runs: 5 datasets $\times$ 4 models). All models received identical instructions; only input placeholders were substituted per run. Design rationale is discussed in Section 3.2 (Table 3).

### A.1 Stage 1 — Detectability Classification Prompt

Applied identically to all four models against the 210-technique input table. Eight calibration examples spanned the full label spectrum: YES (T1071.001, T1190, T1557.001), PARTIAL (T1048.001, T1021.001), and NO (T1059.001, T1040, T1003.001).

| <task> Classify MITRE ATT&CK techniques by NIDS detectability. </task> |
| -------- |
| <role> You are a network security analyst classifying attack techniques |
| based on what a passive network sensor can observe. </role> |
| <sensor_constraints> Passive monitoring only (span port/tap). Observes: |
| IP addresses, ports, protocols, packet headers, unencrypted payloads, |
| flow metadata, timing. Cannot inspect encrypted payloads (TLS/SSL, SSH, |
| VPN) or endpoint activity (process, filesystem, memory, local auth). |
| </sensor_constraints> |
| <detection_methods> 1. SIGNATURE: Pattern matching on headers/payloads |
| 2. ANOMALY: Deviation from behavioral baselines |
| 3. PROTOCOL ANALYSIS: Protocol sequence/state validation |
| 4. FLOW ANALYSIS: Traffic pattern analysis via flow metadata |
| </detection_methods> |
| <classification_labels> |
| YES = Core behavior produces observable network artifacts. |
| PARTIAL = Conditional on encryption state; metadata-only; or variant. |
| NO = Local execution; no network footprint; or traffic |
| indistinguishable from legitimate activity. |
| </classification_labels> |
| <rules> 1. Classify primary mechanism, not follow-on actions. |
| 2. Assume encrypted variant exists where common. |
| 3. If operable without network, classify on local execution. |
| 4. Metadata-only detection (no payload) = PARTIAL. </rules> |
| <output_format> CSV, 5 columns: technique_id, technique_name, |
| Classification, Reasoning (max 15 words), Network_Artifact. |
| Preserve original row order; do not skip or merge rows. </output_format> |
| <examples> [8 calibration examples — see prose above] </examples> |
| <input> {{TECHNIQUE_TABLE}} </input> |

### A.2 Stage 2 — Dataset Coverage Mapping Prompt

Applied 20 times with {{DATASET_NAME}} and {{DATASET_PROFILE}} substituted per run. Each dataset profile supplied to this prompt documented six standardized components: (1) dataset name and collection period, (2) attack categories with record counts and percentage distribution, (3) observable network artifacts per category (ports, protocols, TCP flags, payload characteristics), (4) data format and schema (e.g., Zeek conn.log, PCAP, NetFlow), (5) collection context and traffic generation approach, and (6) known limitations affecting coverage such as class imbalance or missing protocols.

```
| <role> You are a Senior Cyber Threat Intelligence Analyst and NIDS |
| -------- |
| Specialist evaluating coverage of {{DATASET_NAME}} against a fixed |
| list of MITRE ATT&CK techniques. </role> |
| <task> For every technique in the Master List, determine whether any |
| attack scenario in the Dataset Profile corresponds to it. </task> |
| <output_requirements> Format: CSV code block. Rows: Exactly 68 data |
| rows + header (1-to-1, Master List order). Columns: technique_id |
| technique_name | mapped_dataset_label | match_type (Direct/Indirect/ |
| None) | feature_evidence | nist_justification (NIST SP 800-94 terms) |
| </output_requirements> |
| <decision_logic> |
| 1. DIRECT — Explicitly labelled traffic implements core behavior. |
| 2. INDIRECT — Related traffic captures technique as prerequisite, |
| side-effect, or partial artifact. |
| 3. NONE — Neither condition met. |
| </decision_logic> |
| <examples> |
| Direct : T1110 + Credential Access records w/ repeated failed |
| auth flows -> Direct |
| Indirect: T1046 + Reconnaissance scanning implying service |
| enumeration -> Indirect |
| None : T1090 + no C2 obfuscation traffic present -> None |
| </examples> |
| <input> Master List: {{TECHNIQUE_TABLE}} |
| Dataset Profile: {{DATASET_PROFILE}} </input> |
```

**Disclosure of Interests.** The authors have no competing interests to declare that are relevant to the content of this article.

## References

1. Alsaedi, A., Gupta, V., Gupta, B.B.: An overview of ransomware in the financial sector. Journal of Computer Virology and Hacking Techniques **20**, 421–432 (2024), https://doi.org/10.1007/s11416-024-00522-8
2. Anthropic: Use xml tags to structure your prompts. claude documentation (2024), https://docs.anthropic.com/en/docs/use-xml-tags
3. Bagui, S.S., Mink, D., Bagui, S.C., Ghosh, T., Plenkers, R., McElroy, T., Shabanali, S.: Introducing uwf-zeekdata22: A comprehensive network traffic dataset based on the mitre att&ck framework. Data **8**(1), 18 (2023), https://www.mdpi.com/2306-5729/8/1/18
4. Bank for International Settlements: Principles for financial market infrastructures: Executive summary (2023), https://www.bis.org/fsi/fsisummaries/pfmi.htm
5. Bsharat, S., Myrzakhan, A., Shen, Z.: Principled instructions are all you need for questioning llama-1/2, gpt-3.5/4. arXiv preprint arXiv:2312.16171 (2023)

6. Canadian Institute for Cybersecurity, University of New Brunswick: Cic-ids2017 intrusion detection evaluation dataset (2017), https://www.unb.ca/cic/datasets/ids-2017.html
7. Committee on Payments and Market Infrastructures & International Organization of Securities Commissions: Principles for financial market infrastructures (2012), https://www.bis.org/cpmi/publ/d101a.pdf
8. Daniel, N., Kaiser, F.K., Giladi, S., Sharabi, S., Moyal, R., Shpolyansky, S., Puzis, R.: Labeling network intrusion detection system (nids) rules with mitre att&ck techniques: Machine learning vs. large language models. Big Data and Cognitive Computing **9**(2), 23 (2025), https://www.mdpi.com/2504-2289/9/2/23
9. García, S., Grill, M., Stiborek, J., Zunino, A.: An empirical comparison of botnet detection methods. Computers & Security **45**, 100–123 (2014)
10. Hasan, K., Ali, M., Nessa, M., Aditya, S., Mazumder, R.: Retrieval of surface reflectance from noaa-avhrr satellite data. Dhaka University Journal of Engineering and Technology **1**(2), 121–124 (2011)
11. Hasan, K.F., Feng, Y., Tian, Y.C.: Exploring the potential and feasibility of time synchronization using gnss receivers in vehicleto-vehicle communications. In: Proceedings of the 49th Annual Precise Time and Time Interval Systems and Applications Meeting. pp. 80–90 (2018)
12. Hasan, K.F., Shajeeb, H.H., Abeydeera, C., Turnbull, B., Warren, M.: Isadm: An integrated stride, att&ck, and d3fend model for threat modeling against real-world adversaries. IEEE Access **13**, 217316–217348 (2025)
13. Khraisat, A., Gondal, I., Vamplew, P., Kamruzzaman, J.: Survey of intrusion detection systems: techniques, datasets and challenges. Cybersecurity **2**(1), 1–22 (2019), https://link.springer.com/article/10.1186/s42400-019-0038-7
14. Kinnunen, J.: Threat detection gap analysis using mitre att&ck framework (2022), https://www.theseus.fi/handle/10024/745250
15. MITRE: Mitre att&ck v18 (2025), https://attack.mitre.org/
16. Miya, N.F., Joseph, N.: Banking on resilience: 20 years of cybersecurity evolution. South African Journal of Information Management **27**(1), 2019 (2025), https://journals.co.za/doi/full/10.4102/sajim.v27i1.2019
17. Moustafa, N., Slay, J.: Unsw-nb15: a comprehensive data set for network intrusion detection systems (unsw-nb15 network data set). In: 2015 military communications and information systems conference (MilCIS). pp. 1–6. IEEE (2015), https://ieeexplore.ieee.org/abstract/document/7348942
18. Naik, N.: Probabilistic consensus through ensemble validation: A framework for llm reliability. arXiv preprint arXiv:2411.06535 (2024)
19. Rahman Tory, A., Hasan, K.F.: An evaluation framework for network ids/ips datasets: Leveraging mitre att&ck and industry relevance metrics. Computers & Security **153**, 104777 (2025), https://doi.org/10.1016/j.cose.2025.104777
20. Rehman, H.M.R.U., Liaquat, S., Gul, M.J., Jhandir, M.Z., Gavilanes, D., Vergara, M.M., Ashraf, I.: A systematic literature study of machine learning techniques based intrusion detection: datasets, models, challenges, and future directions. Journal of Big Data **12**(1), 264 (2025), https://link.springer.com/article/10.1186/s40537-025-01323-2
21. Scarfone, K., Mell, P.: Guide to intrusion detection and prevention systems (idps). Tech. Rep. NIST SP 800-94, National Institute of Standards and Technology (2007), https://doi.org/10.6028/NIST.SP.800-94
22. Sharafaldin, I., Lashkari, A.H., Hakak, S., Ghorbani, A.A.: Developing realistic distributed denial of service (ddos) attack dataset and taxonomy. In: 2019 International Carnahan Conference on Security Technology (ICCST). IEEE (2019)

23. Tavallaee, M., Bagheri, E., Lu, W., Ghorbani, A.A.: A detailed analysis of the kdd cup 99 data set. In: 2009 IEEE Symposium on Computational Intelligence for Security and Defense Applications. IEEE (2009), https://ieeexplore.ieee.org/abstract/document/5356528
24. Tory, A.R., Hasan, K.F., Rahman, M.S., Koroniotis, N., Moni, M.A.: Mind the gap: Missing cyber threat coverage in nids datasets for the energy sector. arXiv preprint arXiv:2511.00360 (2025), https://doi.org/10.48550/arXiv.2511.00360
25. Wagner, E., Bader, L., Wolsing, K., Serror, M.: Sherlock: A dataset for process-aware intrusion detection research on power grid networks: Dataset paper. In: Proceedings of the Fifteenth ACM Conference on Data and Application Security and Privacy. pp. 419–424 (2024), https://dl.acm.org/doi/abs/10.1145/3714393.3726006
26. Wu, T., Yang, S., Liu, S., Nguyen, D., Jang, S., Abuadbba, A.: Threatmodeling-llm: Automating threat modeling using large language models for banking system. arXiv preprint arXiv:2411.17058 (2024), https://arxiv.org/abs/2411.17058